\documentclass[nopacs,nokeys]{revtex4}%
\usepackage{amsfonts}
\usepackage{amsmath}
\usepackage{amssymb}
\usepackage{bbm}
\usepackage{bm}
\usepackage[utf8]{inputenc}
\usepackage{tensor}
\usepackage{slashed}
\usepackage[centertableaux ]{ytableau}
\usepackage{hyperref}
\usepackage{natbib}
\usepackage{enumerate}
\usepackage{braket,mleftright}
\usepackage{graphicx}
\usepackage[caption=false]{subfig}
\usepackage{subfig}
\usepackage{cleveref}
\usepackage{tcolorbox}
\usepackage{color}
\usepackage{tikz}
\usepackage{tikz,ifthen}%
\providecommand{\U}[1]{\protect\rule{.1in}{.1in}}
\definecolor{jblue}  {RGB}{20,50,100}
  \definecolor{npurple}  {RGB} {153, 51, 204}
  \definecolor{wred}   {RGB}{217,0,56}
  \definecolor{white}   {RGB}{255,255,255}
  \definecolor{korange}   {RGB}{235, 80,  43}
  \definecolor{korange2}   {RGB}{245, 100,  63}
  \definecolor{kyelloworange}   {RGB}{255, 210,  110}
  \definecolor{kyelloworange2}   {RGB}{240, 170,  90}
  \definecolor{kred}   {RGB}{204,  102, 153}
  \definecolor{kpurple}   {RGB}{153,  61, 190}
  \definecolor{kpurplelight}   {RGB}{213,  161, 230}
 \definecolor{nblue}   {RGB}{200,  230, 250}
\definecolor{mycolor}{RGB}{5,61,245}
\usetikzlibrary{trees}
\usetikzlibrary{snakes}
\usetikzlibrary{positioning,decorations.pathmorphing}
\usetikzlibrary{decorations.markings}
\tikzset{beamerprimary/.style={structure.fg, thick}}
\tikzset{beamersecondary/.style={structure.bg, thick}}
\tikzset{ boson/.style={decorate, decoration={snake}},
         gauge/.style={decorate,decoration={snake,post length=1mm}}  ,
     fermion/.style={postaction={decorate},
        decoration={markings,mark=at position .55 with {\arrow{>}}}},
    fermionloop/.style={postaction={decorate},
        decoration={markings,mark=at position .25 with {\arrow{<}}}}, 
    gluon/.style={decorate, 
        decoration={coil,amplitude=4pt, segment length=5pt}},
    scalar/.style={dashed},
    graviton/.style={double}
,
    dm/.style={double,postaction={decorate},decoration={markings,mark=at position .55 with {\arrow{>}}}},
}
\begin{document}
\title{Stationary perturbation theory without sums over intermediate states: Supersymmetric Expansion Algorithm }
\author{M. Napsuciale}
\affiliation{Departamento de F\'{i}sica, Universidad de Guanajuato,
Lomas del Campestre 103, Fraccionamiento
Lomas del Campestre, 37150, Le\'on, Guanajuato, M\'{e}xico.}
\author{S. Rodr\'{\i}guez}
\affiliation{Facultad de Ciencias F\'isico-Matem\'{a}ticas, Universidad Aut\'onoma de Coahuila,
Edificio A, Unidad Camporredondo, 25000, Saltillo, Coahuila, M\'exico.} 

\begin{abstract}
In this work we show that results of Rayleigh-Schr\"{o}dinger perturbation theory can be easily obtained 
using the recently proposed supersymmetric expansion algorithm. Our formalism avoids the sums over 
intermediate states and yield directly corrections to the energy and eigenstates in terms of integrals 
weighted by the probability densities for the edge states of the involved supersymmetric Hamiltonians.

\end{abstract}
\maketitle

Stationary state perturbation theory was developed by Erwin Schr\"{o}dinger in one 
of his seminal papers on quantum mechanics one hundred years ago \cite{Schrodinger:1926vbi}. The formalism, now 
known as Rayleigh-Schr\"{o}dinger perturbation theory, has been widely used in the past and is discussed in most of the textbooks 
on quantum mechanics \cite{Sakurai:2011zz}. The leading correction to the energy is easy to calculate but, 
already at leading order, the corrections to the eigenstates require to calculate an awkward sum over intermediate states. 
Higher order corrections require to consider a new sum over intermediate states every time we increase the order 
of the calculation, which in practice makes the formalism mostly useful for calculations at lowest orders. In the past, alternatives 
have been proposed to deal with the sums over intermediate states in the calculation of corrections 
to the ground state using sum rules \cite{Dalgarno1955}, and other methods \cite{Sternheimer1951}, \cite{Dalgarno1956}
\cite{Aharonov:1979gxt}, \cite{PhysRevA.20.2245}. In particular, Y. Aharonov and C.K. Au use the logarithmic form of 
the Schr\"{o}dinger equation to reduce the calculation of perturbative corrections to the ground state of a one-dimensional 
system to quadrature forms, i.e., to integrals weighted by the ground state probability \cite{Aharonov:1979gxt}, 
avoiding the sums over intermediate states. For excited states, which have nodes, they factorize out the terms containing 
the nodes in the eigenstates, solve for the remaining nodeless part of the eigenstate and update the location of the nodes 
at every order, making the formalism less handy for these states. 

In this letter we show that perturbative calculations to arbitrary order for all the eigenstates can be completely reduced 
to quadrature forms using the recently proposed procedure dubbed supersymmetric expansion algorithm (SEA) 
\cite{Napsuciale:2024yrf}. This procedure avoids the sums over intermediate states of the Rayleigh-Schr\"{o}dinger 
perturbation theory for all the eigenstates, ground or excited. 
The method exploits the hidden supersymmetry of every non-relativistic quantum mechanical problem. The SEA, 
has been recently used to obtain complete analytical solutions to long standing unsolved problemas like the Yukawa potential 
\cite{Napsuciale:2020ehf}, \cite{Napsuciale:2021qtw}, Hulth\'{e}n potential, Anharmonic potential \cite{Napsuciale:2024yrf} 
and Cornell potential \cite{Napsuciale:2025rdr}. All these potentials share the property that the coefficients in the 
expansion of the corresponding physical parameter are polynomials in the respective physical variables, which 
allows us to reduce the Schr\"{o}dinger problem to algebraic recurrence relations. 

An important lesson of the solution of these potentials is that for one-dimensional physical systems the only nodeless 
eigenstate is the ground state. This is well known and it is shown by the solution to the anharmonic potential 
\cite{Napsuciale:2024yrf}. However, for higher dimensional physical systems which can be reduced to a one-dimensional 
effective potential like a central potential, there exist excited nodeless eigenstates. In the case of screening potentials 
like the Yukawa, Hulth\'{e}n and Cornell potentials, there is a nodeless eigenstate for every level $n$, which corresponds 
to the highest value of $l$. These eigenstates are denoted as {\it edge states} and  play a crucial role in solving the problem. 
The edge states are solved first to an arbitrary order in the perturbative expansion using the logarithmic formulation 
of the Schr\"{o}dinger equation. Then, a complete analytical solution for all the eigenstates to arbitrary order can be 
straightforwardly obtained from these nodeless solutions with the aid of supersymmetry. 

In the most general case, the coefficients in the expansion of the potentials are not polynomials and the problem 
cannot be reduced to algebraic recurrence relations. In this work, we show that the SEA can still be used to obtain 
complete analytical solutions to the problem. 
 
 Conventional perturbation theory addresses the problem of a Hamiltonian of the form
 \begin{equation}
 H(\alpha,\beta)= H_{s}(\alpha) + \beta V_{p},
 \label{Hper}
 \end{equation} 
 where we know the solution for the Hamiltonian $H_{s}(\alpha)$ and  $\beta$ is a small parameter controlling the strength 
 of the perturbation. This Hamiltonian has the property
 \begin{equation}
  H(\alpha,\beta) \underset{\beta\to 0}{\longrightarrow} H_{s}(\alpha).
  \label{lim}
 \end{equation}
In this paper we will solve, to arbitrary order in powers of $\beta$, the problem of Hamiltonians satisfying Eq. (\ref{lim}).   
This includes, but does not reduce to, the conventional perturbative problem in Eq. (\ref{Hper}).
 
In order to illustrate the method we consider any physical quantum system which can be reduced to a one-dimension 
Schr\"{o}odinger problem written in terms of dimensionless quantities as 
\begin{equation}
\left[-\frac{d^{2}}{dx^{2}} + v_{0}(x,\lambda)\right]u_{0}(x,\lambda)=\epsilon_{0}(\lambda)u_{0}(x,\lambda),
\label{SEH0}
\end{equation}
with  
\begin{align}
 x=\frac{z}{a},  \quad \lambda=\frac{2\mu a^{2}}{\hbar^{2}} \beta, \quad v_{0}(x,\lambda)=\frac{2\mu a^{2}}{\hbar^{2}} V(a x,\lambda),
  \quad \epsilon_{0}(\lambda)= \frac{2\mu a^{2}}{\hbar^{2}} E(\lambda),
\end{align}
where $z$ denotes the physical variable and $a$ stands for the typical length scale  dictated by the parameter $\alpha$ in 
Eq. (\ref{lim}). It can be a one-dimensional physical system or a higher dimensional system whose solutions can be reduced 
to a one dimensional effective system like the case of central potentials, whose angular angular solutions are given by the spherical 
harmonics and the radial part can be reduced to Eq. (\ref{SEH0}) with an effective potential
\begin{equation}
v_{0}(x,\lambda)\equiv \frac{l(l+1)}{x^{2}} + v(x,\lambda).
\end{equation}
We attach a subscript "$0$" to the dimensionless Hamiltonian 
\begin{equation}
H_{0}(\lambda)=-\frac{d^{2}}{dx^{2}}+v_{0}(x,\lambda),
\end{equation}
and its solutions because this is the first step in the supersymmetric expansion algorithm.

A (nodeless) solution, $u_{0}(x,\lambda)$, to the Schr\"{o}dinger Eq.(\ref{SEH0}), can be obtained from its (non linear) logarithmic form
\begin{equation}
W^{2}_{0}(x,\lambda)-W^{\prime}_{0}(x,\lambda)=v_{0}(x,\lambda) - \epsilon_{0}(\lambda),
\label{LSE0}
\end{equation}
where 
\begin{equation}
W_{0}(x,\lambda)=- \frac{u^{\prime}_{0}(x,\lambda)}{u_{0}(x,\lambda)}=-\frac{d}{dx} \ln (u_{0}(x,\lambda)),
\label{sp0}
\end{equation}
such that the solution is related to the superpotential $W_{0}(x,\lambda)$ as
\begin{equation}
u_{0}(x,\lambda)=e^{-\int W_{0}(x,\lambda)dx}.
\label{u0}
\end{equation}
The solution for the superpotential is obtained expanding in powers of $\lambda$
\begin{equation}
v_{0}(x,\lambda)=\sum_{k=0}^{\infty} v_{0k}(x) \lambda^{k}, \qquad W_{0}(x,\lambda)=\sum_{k=0}^{\infty} w_{0k}(x) \lambda^{k}, 
\qquad \epsilon_{0}(\lambda)=\sum_{k=0}^{\infty}\varepsilon_{0k}\lambda^{k}.
\end{equation}
We reorganize first  the product of the two infinite series defining the non-linear term $W_{0}^{2}$
into a single infinite series to get
\begin{align}
W_{0}^{2}&=\left(\sum^{\infty}_{m=0}w_{0m} \lambda^{m} \right)\left(\sum^{\infty}_{n=0} w_{0n}  \lambda^{n}\right)
=\sum^{\infty}_{m,n=0}w_{0m}w_{0n} \lambda^{m+n} 
\equiv \sum^{\infty}_{k=0} C_{0k} \lambda^{k}.
\end{align}
Here,  the  $C_{0k}$ coefficients are given by the following finite sums,
 \begin{equation}
C_{0k}=\sum_{m+n=k}w_{0m}w_{0n}= \left\{ 
\begin{array}{cc}
w^{2}_{00}, & k=0 \\
2w_{00}w_{0k} + B_{0k}, & k\ge 1 
\end{array}
\right. ,
\end{equation}
with
\begin{equation}
B_{0k}=\sum_{\substack{m+n=k \\ m,n\neq 0 }} w_{0m}w_{0n}.
\end{equation}
It is easy to convince ourselves that $B_{01}=0$ and for $k\ge 2$, $B_{0k}$ involves only $w_{0q}$ with $q=1,2,...,k-1$. 
Using these expressions in the logarithmic Schr\"{o}dinger equation (\ref{LSE0}) we obtain an infinite set of coupled  
first order differential equations
\begin{align}
w_{00}^{2} - w'_{00} &= v_{00} - \varepsilon_{00}, \label{w00eq} \\
2w_{00}w_{0k} -  w^{\prime}_{0k}&= v_{0k} - B_{0k} -  \varepsilon_{0k} , \qquad k\ge 1 \label{w0keq}.
\end{align}

Notice that  this is a cascade of hierarchical equations where now only the $k=0$ case is still non-linear. However, the
$k=0$ case corresponds to the unperturbed system whose solutions are known. The remaining equations are linear and can be 
solved as described below. In the case when $v_{0k}$ are polynomials, the solutions to these equations are 
also polynomials which allows us to reduce the set of linear differential equations to algebraic recurrence relations for the 
coefficients of the polynomial solutions which can be solved to arbitrary order as done in 
\cite{Napsuciale:2020ehf},\cite{Napsuciale:2021qtw}, for the Yukawa potential, in \cite{Napsuciale:2024yrf} for the 
Hulth\'{e}n and anharmonic potential, and in \cite{Napsuciale:2025rdr} for the Cornell potential.

In the most general case, Eqs. (\ref{w00eq}, \ref{w0keq}) do not admit  polynomial solutions. However, the supersymmetric expansion 
algorithm still can be used to obtain a solution as follows. To leading order, Eq. (\ref{u0}) yields
\begin{equation}
u_{0}(x,0)=e^{-\int w_{00}(x) dx} .
\label{u00}
\end{equation}
This solution is known, thus the leading term in the superpotential is given by
\begin{equation}
w_{00}(x) = - \frac{u^{\prime}_{0}(x,0)}{u_{0}(x,0) }.
\label{w00u0}
\end{equation} 

The corrections in the power expansion in $\lambda$ are obtained from Eq. (\ref{w0keq}). Multiplying by $u^{2}_{0}(x,0)$ this equation and 
using Eq. (\ref{w00u0}) we get
\begin{align}
-\frac{d}{dx}\left(u^{2}_{0}(x,0)w_{0k}(x)\right) &=(v_{0k}(x) -B_{0k}(x) -\varepsilon_{0k})u^{2}_{0}(x,0).
\end{align}
Integrating this equation over the whole space we obtain
\begin{align}
 \varepsilon_{0k} =\langle u_{0}(x,0)| v_{0k}(x)-B_{0k}(x)|u_{0}(x,0)\rangle.
 \label{e0k}
\end{align}
Integrating instead partially from the minimal admissible value $x_0$ to $x$ we get
\begin{align}
 w_{0k} (x) &=\frac{1}{u^{2}_{0}(x,0)}\int_{x_0}^{x}(\varepsilon_{0k}-v_{0k}(t) + B_{0k}(t) )u^{2}_{0}(t,0)dt.
 \label{w0k}
 \end{align}
 Notice that for every potential $B_{01}=0$ thus, for perturbative calculations, Eq. (\ref{e0k}) with $k=1$ yields the 
 conventional leading result of Rayleigh-Schr\"odinger perturbation theory. However, the leading correction to the 
 first nodeless eigenstate of $H_{0}$ and higher order corrections are given in terms of integrals weighted by 
 the unperturbed probability density $u^{2}_{0}(x,0)$, i.e., in quadrature forms, instead of sums over intermediate 
 states. The hierarchy of Eqs. (\ref{w00eq},\ref{w0keq}) allows us to easily 
 calculate $u_{0}(x,\lambda)$ to the desired order in $\lambda$ starting with $k=0$. 
At a given order $k$ we know already the solutions to order $k-1$ which are needed to calculate the corrections to order $k$. 

Results in Eqs. (\ref{e0k},\ref{w0k} are similar to those obtained in \cite{Aharonov:1979gxt} 
for one-dimensional physical systems where the only nodeless eigenstate is the ground state, but the use of 
superpotentials makes the procedure more transparent. Furthermore, for states with nodes (excited states 
in the case on one-dimensional physical systems) it connects in a natural way with supersymmetry which can 
be used to solve these states as described in the following.

Once we have solved the nodeless eigenstate $u_{0}(x,\lambda)$, we solve for eigenstates with one node as follows. 
First, the Hamiltonian $H_{0}$ can be factorized as
\begin{equation}
H_{0}=a^{\dagger}_{0} a +\epsilon_{0},
\end{equation}
where
\begin{equation}
a^{\dagger}_{0}= - \frac{d}{dx} + W_{0}(x,\lambda), \qquad  a_{0}=\frac{d}{dx} + W_{0}(x,\lambda).
\end{equation}
The supersymmetric partner Hamiltonian 
\begin{equation}
H_{1}=a_{0}a_{0}^{\dagger} + \epsilon_{0}(\lambda)= -\frac{d^{2}}{dx^{2}}+ v_{1}(x,\lambda),
\end{equation}  
has the following potential
\begin{align}
 v_{1}(x,\lambda)&= v_{0}(x,\lambda) + 2 W_{0}^{\prime}(x,\lambda) =\sum_{k=0}^{\infty} v_{1k}(x) \lambda^{k},
\end{align}
and has the same spectrum as $H_{0}$ except for the $\epsilon_{0}(\lambda)$ level \cite{Witten:1981nf}.

We solve likewise the Schr\"{o}dinger equation for the nodeless eigenstate of the superpartner $H_{1}$
\begin{equation}
 H_{1}(\lambda)u_{1}(x,\lambda)= \left[ -\frac{d^{2}}{dx^{2}} +v_{1}(x,\lambda)  \right]u_{1}(x,\lambda)
 = \epsilon_{1}(\lambda) u_{1} (x,\lambda).
 \label{SE1}
 \end{equation}
Casting this equation into its logarithmic form we get
\begin{equation}
W^{2}_{1}(x,\lambda)-W^{\prime}_{1}(x,\lambda)=v_{1}(x,\lambda) - \epsilon_{1}(\lambda),
\end{equation}
where 
\begin{equation}
W_{1}(x,\lambda)=-\frac{u^{\prime}_{1}(x,\lambda)}{u_{1}(x,\lambda)}=-\frac{d}{dx} \ln (u_{1}(x,\lambda)),
\label{sp1}
\end{equation}
and expanding in powers of $\lambda$
\begin{equation}
W_{1}(x,\lambda)=\sum_{k=0}^{\infty} w_{1k}(x) \lambda^{k}, \qquad \epsilon_{1}(\lambda)=\sum_{k=0}^{\infty}\varepsilon_{1k}\lambda^{k},
\end{equation}
we get the analogous set of hierarchical equations
\begin{align}
w_{10}^{2} - w'_{10} &= v_{10}(x) - \varepsilon_{10}, \label{w10eqgen} \\
2w_{10}w_{1k} -  w^{\prime}_{1k}&=v_{1k}(x) - B_{1k} -  \varepsilon_{1k} , \qquad k\ge 1 \label{w1keqgen},
\end{align}
where
\begin{equation}
B_{1k}=\sum_{\substack{m+n=k \\ m,n\neq 0 }} w_{1m}w_{1n}.
\end{equation}

The nodeless solution for $H_{1}$ is given by
\begin{equation}
u_{1}(x,\lambda)=e^{-\int W_{1}(x,\lambda) dx},
\label{u1}
\end{equation}
and the corresponding one-node solution for $H_{0}$, with the common energy $\epsilon_{1}(\lambda)$, is given by
\begin{equation}
u_{10}(x,\lambda)=a^{\dagger}_{0}u_{1}(x,\lambda)=\left( -\frac{d}{dx}+W_{0}(x,\lambda) \right) u_{1}(x,\lambda).
\label{u10sol}
\end{equation}
At this point, the main problem is to solve the non-linear leading Eq.(\ref{w10eqgen}). Somehow, the solution to this equation 
must be related to the solutions of the unperturbed problem. The link is stablished considering Eq. (\ref{u10sol}) to leading 
order ($\lambda=0$) which yields
\begin{equation}
u_{10}(x,0)=\left( -\frac{d}{dx}+w_{00}(x) \right) u_{1}(x,0),
\label{u10eq}
\end{equation}
where, according to Eq. (\ref{u1})
\begin{equation}
u_{1}(x,0)=e^{-\int w_{10}(x) dx}.
\label{u1zero}
\end{equation}
But we know the solutions of $H_{0}$ for $\lambda=0$ (the unperturbed system in the case of perturbation theory),
thus we know $u_{10}(x,0)$ as well as the corresponding leading term in the expansion of the energy, 
$\varepsilon_{10}$. We can use this information 
to solve instead Eq. (\ref{u10eq}) for $u_{1}(x,0)$ to obtain later $w_{10}$ from this solution. Indeed, multiplying 
Eq. (\ref{u10eq}) by $u_{0}(x,0)$ after simple manipulations we get
\begin{align}
  u_{0}(x,0) u_{10}(x,0)= - \frac{d }{dx}( u_{0}(x,0)u_{1}(x,0)),
\end{align}
which yields the nodeless solution of $H_{1}$ for $\lambda=0$ as
\begin{equation}
u_{1}(x,0)= - \frac{1}{u_{0}(x,0)} \int_{x_{0}}^{x} u_{0}(t,0) u_{10}(t,0) dt.
\label{u1sol0}
\end{equation}
From this solution we get the following solution to Eq. (\ref{w10eqgen})
\begin{equation}
w_{10}(x)=- \frac{u^{\prime}_{1}(x,0)}{u_{1}(x,0)}.
\end{equation}

The perturbative corrections for the nodeless eigenstates of $H_{1}$ are obtained from Eq. (\ref{w1keqgen}) 
in a similar way as we did for the nodeless eigenstates of $H_{0}$. We get the analogous results
 \begin{align}
 \varepsilon_{1k} &= \langle u_{1}(x,0)|   v_{1k}(x) - B_{1k}(x)|u_{1}(x,0)\rangle, \label{e1k}\\
 w_{1k} (x) &=\frac{1}{u^{2}_{1}(x,0)}\int_{0}^{x}(\varepsilon_{1k}+B_{1k}(t)-v_{1k}(t) )u^{2}_{1}(t,0)dt. \label{w1k}
\end{align}
These results yield the nodeless eigenstates of $H_{1}$ to arbitrary order, which in turn yields the one-node solutions 
of $H_{0}$ through Eq. (\ref{u10sol}). The corresponding eigenvalue is $\epsilon_{1}(\lambda)$ which is shared by 
$H_{0}$ and $H_{1}$.

The eigenstates with $r$ nodes are obtained similarly with the aid of supersymmetry. 
The $r$-th supersymmetric partner, $H_{r}$, has the potential
\begin{equation}
v_{r}(x,\lambda)=v_{0}(x,\lambda)+2\sum_{q=0}^{r-1}W^{\prime}_{q}(x,\lambda).
\end{equation}
The nodeless solution for $H_{r}$ is obtained from the logarithmic form of the SE 
\begin{equation}
W^{2}_{r}(x,\lambda) - W^{\prime}_{r}(x,\lambda)=v_{r}(x,\lambda)-\epsilon_{r}(\lambda),
\end{equation}
as
\begin{equation}
u_{r}(x,\lambda)=e^{-\int W_{r}(x,\lambda)} dx.
\end{equation}
The superpotential $W_{r}(x,\lambda)$ is obtained expanding in powers of $\lambda$
\begin{align}
v_{r}(x,\lambda)=\sum_{k=0}^{\infty}v_{rk}(x) \lambda^{k}, \qquad W_{r}(x,\lambda)=\sum_{k=0}^{\infty}w_{rk}(x) \lambda^{k}, 
\qquad  \epsilon_{r}(\lambda)=\sum_{k=0}^{\infty} \varepsilon_{rk}\lambda^{k}.
\end{align}
This expansion yields the infinite cascade of coupled first order differential equations
\begin{align}
w_{r0}^{2} - w'_{r0} &= v_{r0}(x) - \varepsilon_{r0}, \label{wr0} \\
2w_{r0}w_{rk} -  w^{\prime}_{rk}&=  v_{rk}- B_{rk}  -  \varepsilon_{rk} , \qquad k\ge1. \label{wrk}
\end{align}
The leading equation (\ref{wr0}) is non-linear but it can be solved using the information of the unperturbed system 
as shown below, which yields the leading term $u_{r}(x,0)$. The corrections to order $\lambda^{k}$ are then obtained as
\begin{align}
 \varepsilon_{rk} &= \langle u_{r}(x,0)| v_{rk}(x) - B_{rk}(x)|u_{r}(x,0)\rangle, \label{erkgen}\\
 w_{rk} (x) &=\frac{1}{u^{2}_{r}(x,0)}\int_{x_{min}}^{x}(\varepsilon_{rk} - v_{rk}(t) + B_{rk}(t) )u^{2}_{r}(t,0)dt. \label{wrkgen}
\end{align}

The eigenvalue $\epsilon_{r}(\lambda)$ corresponding to the nodeless solution of $H_{r}$ is a common eigenvalue of the 
set $\{ H_{0}, H_{1},..., H_{r} \}$ with the eigenstate of $H_{q}$ given by $u_{r,q}(x,\lambda)$, $q=r-1, r-2, ...,0$ where
\begin{align}
u_{r,r-1}(x,\lambda)&= a^{\dagger}_{r-1}u_{r}(x,\lambda)= \left( -\frac{d}{dx}+W_{r-1}(x,\lambda) \right)u_{r}(x,\lambda), 
\label{urrm10}\\
u_{r,r-2}(x,\lambda)&= a^{\dagger}_{r-2}u_{r,r-1}(x,\lambda)= \left( -\frac{d}{dx}+W_{r-2}(x,\lambda) \right)u_{r,r-1}(x,\lambda), 
\label{urrm20} \\
.....& ....... \nonumber \\
u_{r,0}(x,\lambda)&= a^{\dagger}_{0}u_{r,1}(x,\lambda)= \left( -\frac{d}{dx}+W_{0}(x,\lambda) \right)u_{r,1}(x,\lambda). 
\label{ur00}
\end{align}

The main problem in the $r$-th step is to solve Eq. (\ref{wr0}). But we know the solution of $H_{0}$ for $\lambda=0$, 
the unperturbed solution $u_{r,0}(x,0)$, which can be used for this purpose. Indeed, for $\lambda=0$ equations 
(\ref{urrm10},\ref{urrm20},\ref{ur00}) read
\begin{align}
u_{r,r-1}(x,0)&= a^{\dagger}_{r-1}u_{r}(x,0)= \left( -\frac{d}{dx}+w_{r-1,0}(x) \right)u_{r}(x,0), \\
u_{r,r-2}(x,0)&= a^{\dagger}_{r-2}u_{r,r-1}(x,0)= \left( -\frac{d}{dx}+w_{r-2,0}(x) \right)u_{r,r-1}(x,0),  \\
.....& ....... \nonumber\\
u_{r,0}(x,0)&= a^{\dagger}_{0}u_{r1}(x,0)= \left( -\frac{d}{dx}+w_{00}(x) \right)u_{r1}(x,0). 
\end{align}
We solve first the last equation for $u_{r1}(x,0)$ and then we solve recursively the previous equations to obtain
\begin{align}
u_{r,1}(x,0)&=-\frac{1}{u_{0}(x,0)} \int u_{0}(x,0) u_{r,0}(x,0) dx, \label{ur10sol} \\
u_{r,2}(x,0)&=-\frac{1}{u_{1}(x,0)} \int u_{1}(x,0) u_{r,1}(x,0) dx, \label{ur20sol}\\
.....& .....\nonumber \\
u_{r}(x,0)&=-\frac{1}{u_{r-1}(x,0)} \int u_{r-1}(x,0) u_{r,r-1}(x,0) dx. \label{ur00sol}
\end{align}
The solution to Eq.(\ref{wr0}) is given by
\begin{equation}
w_{r0}=- \frac{d}{dx} \ln (u_{r}(x,0)),
\end{equation}
which can then be used to solve the problem to arbitrary order using Eqs. (\ref{erkgen},\ref{wrkgen}).

The main results of this work are given by the general relations in Eqs. Eqs. (\ref{erkgen},\ref{wrkgen}) and the 
connection with the information of the known solutions of $H_{0}$ in Eqs. (\ref{ur10sol}, \ref{ur20sol}, \ref{ur00sol}).
The formalism yields the complete analytical solution of the perturbed problem in the form of a power series in $\lambda$. 
As a cross check we used it to solve the Yukawa, Hulth\'{e}n, anharmonic and Cornell potentials, which have polynomial 
solutions for the superpotentials, obtaining the same results as the algebraic recursive relations 
given in \cite{Napsuciale:2024yrf}, \cite{Napsuciale:2020ehf}, \cite{Napsuciale:2021qtw} \cite{Napsuciale:2025rdr}.
In the case of perturbation theory, the expansion parameter $\lambda$ is small and we can keep a few terms according 
to the precision required by experimental data.

As an example of a one-dimensional physical system with non-polynomial superpotentials we consider the 
perturbative corrections to order $\lambda^{3}$  of a particle in a box of width $a$ which has the potential 
\begin{equation}
V(z) =
\left\{ \begin{array}{cc} 
\infty, & \quad |z|\geq \frac{a}{2},\\  
0, &\quad |z|< \frac{a}{2}.\\
 \end{array} \right.
\end{equation}
due to a harmonic oscillator perturbation 
\begin{equation}
\beta V_{p}(z)=  \frac{1}{2}\mu \omega^{2} z^{2}.
\end{equation}
In this case, the natural scale length is the box width $a$ and the dimensionless quantities are given by
 \begin{align}
 x=\frac{z}{a},  \quad \lambda=\frac{\mu^{2} \omega^{2} a^{4}}{\hbar^{2}} , 
 \quad v_{0}(x,\lambda)=\frac{2\mu a^{2}}{\hbar^{2}} V(a x,\lambda),
  \quad \epsilon_{0}(\lambda)= \frac{2\mu a^{2}}{\hbar^{2}} E(\lambda).
 \end{align}
The dimensionless Schr\"{o}dinger equation reads
 \begin{equation}
 H_{0}(\lambda)u_{0}(x,\lambda)\equiv \left[ -\frac{d^{2}}{dx^{2}} +v_{0}(x,\lambda)  \right]u_{0}(x,\lambda)
 = \epsilon_{0}(\lambda) u_{0} (x,\lambda),
 \label{SEB}
 \end{equation} 
with 
\begin{equation}
v_{0}(x,\lambda) =
\left\{ \begin{array}{cc} 
\infty, & \quad |x|\geq \frac{1}{2},\\  
\lambda x^{2}, &\quad |x| < \frac{1}{2}.\\
 \end{array} \right.
\end{equation}

The solutions for the eigenfunctions are nule outside the box and the logarithmic SE for the leading term of the superpotential 
inner region is
\begin{align}
w_{00}^{2} - w'_{00} &= - \varepsilon_{00}. 
\end{align}
The general solution to this equation is
\begin{equation}
w_{00}(x)= k tan (kx), \qquad \varepsilon_{00}=k^{2},
\end{equation}
with arbitrary values of $k$. However, boundary conditions requires the wave function
\begin{equation}
u_{0}(x,0)=Exp\left[-\int w_{00}(x) dx\right]= Exp\left[-\int \frac{k \sin(kx)dx}{\cos(kx) }\right]=Exp\left[\ln(\cos(kx))+c\right]=N \cos(kx),
\end{equation}
to vanish at $x=1/2$. On the other hand it must be a nodeless function. The first condition yields $k/2=(n+1/2)\pi$, with $n=0,1,2,...$.
The second condition restricts the solution obtained by the logarithmic SE to $k=\pi$, thus, the leading terms of the expansions of 
the superpotential and the energy for the nodeless solution of $H_{0}$ are given by
\begin{equation}
w_{00}(x)= \pi \tan (\pi x), \qquad \varepsilon_{00}=\pi^{2},
\end{equation}
which correspond to the ground state of the unperturbed system. For this one-dimensional system there exist only one 
nodeless eigenstate which corresponds to the ground state. The normalized eigensolution for the unperturbed ground state 
is obtained from Eq. (\ref{u00}) as
 \begin{align}
u_{0}(x,0)=\left\{ \begin{array}{cc} 
0, & \quad |x|\geq \frac{1}{2},\\  
\sqrt{2} \cos(\pi x), &\quad |x| < \frac{1}{2}.\\
 \end{array} \right. 
\end{align}

Straightforward calculations using Eqs. (\ref{erkgen},\ref{wrkgen}) yield the solution to the desired 
order in $\lambda$. We obtain long expressions for the superpotentials and the eigenstates which are not enlightening and 
quote here just the results for the eigenvalues to order $\lambda^{3}$ for the gound and first excited states. 
The energy of the ground state to order $\lambda^{3}$ is given by
\begin{align}
\epsilon_{0}(\lambda) &= \pi^{2} + \frac{\pi^{2}-6}{12 \pi^2}  \lambda + \frac{630-75 \pi ^2+\pi ^4}{720 \pi^6} \lambda^{2}
+\frac{-228690+29295 \pi ^2-630 \pi^4+\pi ^6}{30240 \pi ^{10}} \lambda^{3}.  
\end{align}

The calculation of the perturbative corrections to the first excited state requires to calculate first the nodeless 
solution of the first unperturbed supersymmetric Hamiltonian $H_{1}$, i.e., for $|x|<1/2$, to find the solution to
\begin{align}
w_{10}^{2} - w'_{10} & = v_{0}(x) + 2w^{\prime}_{00}(x) - \varepsilon_{10} = 2\pi^{2}\sec^{2}(\pi x)- \varepsilon_{10}.
\label{w10eq}
\end{align}
The solution for the first excited state of $H_{0}$ for $\lambda=0$ in the inner region is given by
\begin{align}
u_{10}(x,0)= \sqrt{2} \sin(2\pi x), \qquad \varepsilon_{10}=4\pi^{2}.
\end{align}
The normalized nodeless solution of $H_{1}$ to leading order is obtained from Eq. (\ref{u1sol0}) as
\begin{align}
u_{1}(x,0)=\sqrt{\frac{8}{3}} \cos^{2}(\pi x).
\end{align}
which yields the following solution to Eq. (\ref{w10eq})
\begin{equation}
w_{10}(x)=- \frac{u^{\prime}_{1}(x,0)}{u_{1}(x,0)}= 2\pi \tan (\pi x).
\end{equation}

The energy of the first excited state to order $\lambda^{3}$ is straightforwardly obtained  
from Eqs. (\ref{erkgen},\ref{wrkgen}) as
\begin{align}
\varepsilon_{1}(\lambda) &= 4\pi^{2}+ \frac{2 \pi ^2-3}{24 \pi ^2}\lambda 
+ \frac{315-150 \pi ^2+8 \pi^4}{23040 \pi ^6} \lambda^{2}
+ \frac{-114345+58590 \pi ^2-5040 \pi^4+32 \pi ^6}{15482880 \pi^{10}} \lambda^{3}.
 \end{align}  
 
\bibliography{PTWT}
\bibliographystyle{JHEP}
\end{document}